\documentclass[usenatbib,usedcolumn,usegraphicx,fleqn]{mn2e}
\pdfoutput=1
\usepackage[totalwidth=480pt,totalheight=680pt]{geometry}
\usepackage{amsmath}
\usepackage{amssymb}
\usepackage{color}
\usepackage{ifthen}

\title[\lae{} clustering around luminous quasars]{Clustering of $\bmath{\lya{}}$ emitters around luminous quasars at $\bmath{z=2}$--$\bmath{3}$: an alternative probe of reionization on galaxy formation} 
\author[Bruns Jr. et al.]{Loren R. Bruns Jr.$^1$\thanks{Email: lbrunsjr@student.unimelb.edu.au}, J. Stuart B. Wyithe$^1$\thanks{Email: swyithe@unimelb.edu.au}, Joss Bland-Hawthorn$^2$\thanks{Email: jbh@physics.usyd.edu.au},
\newauthor and Mark Dijkstra$^3$\thanks{Email: dijkstra@MPA-Garching.MPG.DE} \\
$^1$School of Physics, University of Melbourne, Parkville, Victoria 3010, Australia \\
$^2$Sydney Institute for Astronomy, School of Physics, University of Sydney, NSW 2006, Australia \\
$^3$Max-Planck-Institut f\"{u}r Astrophysik, Karl-Schwarzschild Str.\,1, 85748 Garching, Germany}

\providecommand{\modecheck}[2][]{\ifthenelse{\equal{#1}{}}{\ifmmode#2\else$#2$\fi}{\ifmmode#2\else#1\fi}}
\providecommand{\sub}[2]{\modecheck{{#1}_{\scriptscriptstyle \rmn{#2}}}}

\providecommand{\metalline}[2]{\modecheck[\mbox{#1\,{\sc #2}}]{\rmn{#1}\,\textsc{#2}}}

\providecommand{\angstrom}{\modecheck[\AA]{\rmn{\text{\AA}}}}
\providecommand{\centimeter}{\modecheck{\rmn{cm}}}

\providecommand{\cmpc}{\modecheck{\rmn{cMpc}}}
\providecommand{\ergs}{\modecheck{\rmn{ergs}}}
\providecommand{\hertz}{\modecheck{\rmn{Hz}}}
\providecommand{\kelvin}{\modecheck{\rmn{K}}}

\providecommand{\lbsolar}{\sub{\rmn{L}}{\sun, B}}
\providecommand{\logten}{\sub{\log}{10}}
\providecommand{\msolar}{\sub{\rmn{M}}{\sun}}
\providecommand{\planck}{\sub{h}{\rmn{p}}}
\providecommand{\pkpc}{\modecheck{\rmn{pkpc}}}
\providecommand{\pmpc}{\modecheck{\rmn{pMpc}}}
\providecommand{\second}{\modecheck{\rmn{s}}}
\providecommand{\steradian}{\modecheck{\rmn{sr}}}
\providecommand{\yr}{\modecheck{\rmn{yr}}}
\providecommand{\Zsolar}{\modecheck{\rmn{Z}_{\sun}}}

\providecommand{\igm}{\modecheck[IGM]{\rmn{IGM}}}

\providecommand{\lae}{\modecheck[LAE]{\rmn{LAE}}}

\providecommand{\lya}{\modecheck[Ly$\alpha$]{\rmn{Ly}\alpha}}

\providecommand{\jossquasar}{\modecheck[PKS 0424-131]{\rmn{PKS 0424-131}}}

\providecommand{\dutycycle}{\sub{\epsilon}{\rmn{DC}}}
\providecommand{\equivwidth}{\modecheck{\rmn{EW}}}
\providecommand{\fdilution}{\sub{f}{\rmn{dil}}}
\providecommand{\fdust}{\sub{f}{\rmn{dust}}}
\providecommand{\fescape}{\sub{f}{\rmn{esc}}}
\providecommand{\fstar}{\sub{f}{\star}}
\providecommand{\hubble}{\modecheck{h}}
\providecommand{\hubbletime}{\sub{t}{\rmn{hub}}}
\providecommand{\igmtrans}{\sub{\mathcal{T}}{\igm{}}}
\providecommand{\ionlum}{\sub{\dot{Q}}{\rmn{H}}}
\providecommand{\lyalum}{\sub{L}{\lya{}}}
\providecommand{\mfp}{\sub{r}{mfp}}
\providecommand{\nfrac}{\sub{\chi}{\igm{}}}
\providecommand{\omegab}{\sub{\Omega}{b}}
\providecommand{\omegal}{\sub{\Omega}{\Lambda}}
\providecommand{\omegam}{\sub{\Omega}{m}}
\providecommand{\sigmaeight}{\sub{\sigma}{8}}
\providecommand{\sfr}{\sub{\dot{M}}{\star}}
\providecommand{\tilt}{\sub{n}{s}}
\providecommand{\virialtemp}{\sub{T}{\rmn{vir}}}
\providecommand{\uvbg}{\sub{\Gamma}{BG}}
\providecommand{\uvcontlum}{\sub{L}{\lambda, UV}}
\providecommand{\wltrans}{\modecheck{\langle \rmn{e}^{-\tau}\rangle}}

\providecommand{\lyafreq}{\sub{\nu}{\lya{}}}
\providecommand{\lyawl}{\sub{\lambda}{\lya{}}}

\providecommand{\blackbold}{\modecheck{\textcolor{black}{\textbf{black}}}}
\providecommand{\redbold}{\modecheck{\textcolor{red}{\textbf{red}}}}

\providecommand{\eqnref}[1]{Eqn.\,(\ref{#1})} 
\providecommand{\figref}[1]{Fig.\,\ref{#1}} 
\providecommand{\secref}[1]{\S\ref{#1}}
\providecommand{\tblref}[1]{Table\,\ref{#1}}

\defcitealias{Francis:2004p2989}{FBH04}
\defcitealias{Ouchi:2008p3716}{O08}
\defcitealias{Dijkstra:2007p1}{D07}

\begin{document}
\maketitle

\begin{abstract}
Narrowband observations have detected no \lya{} emission within a $70\,\pmpc{}^{3}$ volume centered on the $z=2.168$ quasar \jossquasar{}. This is in contrast to surveys of \lya{} emitters in the field at similar redshifts and flux limits, which indicate that tens of sources should be visible within the same volume.  The observed difference indicates that the quasar environment has a significant influence on the observed density of \lya{} emitters. To quantify this effect we have constructed a semi-analytic model to simulate the effect of a luminous quasar on nearby \lya{} emitters. We find the null detection around \jossquasar{} implies that the minimum isothermal temperature of \lya{} emitter host halos is greater than $3.4\times10^{6}\,\kelvin{}$ ($68\%$ level), corresponding to a virial mass of $\sim1.2\times10^{12}\,\msolar{}$. This indicates that the intense UV emission of the quasar may be suppressing the star formation in nearby galaxies.  Our study illustrates that low redshift quasar environments may serve as a surrogate for studying the radiative suppression of galaxy formation during the epoch of reionization.
\end{abstract}


\section[Introduction]{Introduction}
Current models of reionization are constrained by observation to begin at $z\sim10$ \citep{Komatsu:2011in} and to have been completed by $z\sim6$ \citep{Fan:2006p777}.  The popular picture for this process assumes that isolated and internally ionized ultraviolet (UV) sources carved out bubbles of ionized hydrogen (\metalline{H}{ii}) in the neutral intergalactic medium (\igm{}).  These bubbles grew in size and increased in number as the cosmic star formation rate increased and more UV sources illuminated the \igm{}.  Eventually these bubbles overlapped until they pervaded all of space, leaving the entire \igm{} ionized and thus ending reionization \citep[for a review on reionization see][]{Barkana:2001p1219}. The higher \igm{} temperature in these ionized regions raised the isothermal virial temperature required for gas accretion onto a dark matter halo \citep{Dijkstra:2004p5194} greatly increasing the critical mass required to form galaxies.  This process of raising the minimum halo mass for galaxy formation -- known as `Jeans-mass filtering' -- is thought to have played a crucial role in the transition to an ionized \igm{} \citep{Gnedin:2000p5990}. Learning how this mechanism works is therefore vital to our understanding of reionization \citep{Iliev:2005p5296}.

Observing Jeans-mass filtering during the epoch of reionization directly is not possible at present, making it difficult to determine its role with respect to completing reionization within the observed timeline. To constrain different mechanisms of galaxy formation during reionization with current instruments therefore requires a surrogate environment that can be readily observed, such as the dense and ionized regions around quasars.  By collecting statistics about the number densities and masses of galaxies within the highly ionized and clustered regions around quasars we can get an observational handle on the effects of a highly ionized environment on these galaxies.  The goal of this paper is to compare the observed number density of galaxies with a theoretical model in order to highlight any environmental effects introduced by the highly ionized environment.

\citet[][hereafter \citetalias{Francis:2004p2989}]{Francis:2004p2989} imaged the region around the luminous $z=2.168$ quasar \jossquasar{}, looking for fluorescent \lya{} emission from clouds of neutral hydrogen (\metalline{H}{i}).  They used the Taurus Tunable Filter \citep{BlandHawthorn:1998tx} on the Anglo-Australian Telescope to probe a volume of $70\,\pmpc{}^{3}$ centered on the quasar with three narrow-band ($7\,\angstrom{}$ FWHM) images tuned to rest frame \lyawl{} at $z=2.161$, $z=2.168$, and $z=2.175$.  This technique provides low resolution spectra across the entire field of view, allowing any source of \lya{} emission to be easily selected by looking for dropouts in the three redshift bands.  Based on surveys done at similar redshifts and accounting for galaxy clustering, \citetalias{Francis:2004p2989} expected to see between $6$ and $25$ fluorescing hydrogen clouds of varying sizes and $\gtrsim{}10$ internally ionized \lya{} emitting galaxies (\lae{}s).  However, their observations found no hydrogen clouds, nor any \lae{}s, leading them to the tentative conclusion that quasar induced photo-evaporation was destroying the clouds and preventing or suppressing the formation of stars in nearby galaxies.

In this paper we construct a semi-analytic model to interpret the observation in \citetalias{Francis:2004p2989}.  A null detection of \lae{}s in these regions either means that galaxies do not exist near to the quasar (i.e. were destroyed or never formed) or that they are not detectable (their emission is obscured by dense patches of the \igm{} or by interstellar dust) at the epoch of observation. Our semi-analytic model includes the effects of \lya{} transmission and galaxy clustering, and can be used to explore the implications of radiative feedback on galaxy formation.  We tailor this model to the environment of a luminous quasar, and populate it with star forming galaxies following a density prescription that is calibrated against luminosity functions from wide-field \lae{} surveys \citep[][hereafter \citetalias{Ouchi:2008p3716}]{Ouchi:2008p3716}.  

We describe the model in \secref{sec:modeling}, and present comparison with observations in \secref{sec:comparison}, along with a discussion on how consistent a null detection is with this model.  In \secref{sec:conclusions} we summarize our findings, argue that there is a strong indication of the suppression of low mass galaxies within in the volume around \jossquasar{}, and suggest that further surveys of \lae{}s near quasars can yield constraints on the process of Jeans-mass filtering during reionization.  We assume the standard WMAP7 cosmology \citep{Komatsu:2011in}, (\omegam{}, \omegal{}, \omegab{}, \hubble{}, \sigmaeight{}, \tilt{}) = ($0.27$, $0.73$, $0.046$, $0.70$, $0.81$, $0.96$).  Throughout we adopt the convention of specifying a distance as physical or co-moving by prepending a `p' or `c' to the distance unit (i.e. \pmpc{} and \cmpc{}).

\section[Modeling of \texorpdfstring{\lya{}}{Ly-alpha} galaxies in quasar environments]{Modeling of \texorpdfstring{$\bmath{\lya{}}$}{Ly-alpha} galaxies in quasar environments}
\label{sec:modeling}
This section outlines the model used in our analysis. The model is divided into two parts, i) a calculation of the observed \lya{} luminosity and number density for galaxies of a particular mass, and ii) how these relationships are affected by the exotic environment of a nearby quasar.  The observed \lya{} luminosity is determined by the intrinsic \lya{} luminosity (\secref{sec:lya_lum}) multiplied by the fraction of this luminosity that is transmitted through the \igm{} (\secref{sec:igm_transmission}). The model also yields a UV luminosity (\secref{sec:uv_lum}) allowing us to fit both \lya{} and UV luminosity functions to determine the free parameters in the model (\secref{sec:luminosity_functions}). Finally, the model also accounts for the impact of the quasar on the \igm{} near the galaxies (\secref{sec:quasar_transmission}).

\subsection[Galactic \texorpdfstring{\lya{}}{Ly-alpha} Luminosity]{Galactic \texorpdfstring{$\bmath{\lya{}}$}{Ly-alpha} Luminosity}
\label{sec:lya_lum}
To determine the intrinsic \lya{} luminosity corresponding to a given halo mass we follow the semi-analytic model outlined in \citet[][hereafter \citetalias{Dijkstra:2007p1}]{Dijkstra:2007p1}.  This model first determines the galaxy's star formation rate (\sfr{}) then uses this to calculate the resultant ionizing luminosity (\ionlum{}) and hence a \lya{} luminosity (\lyalum{}).  These relationships depend on the unitless free parameters for duty cycle (lifetime as a fraction of the Hubble time; \dutycycle{}), star formation efficiency (\fstar{}), and ionizing escape fraction (\fescape{}).

For a halo with mass $m$ at redshift $z$ we define the star formation rate to be:
\begin{equation}
  \label{eqn:star_formation_rate}
  \sfr{}(m, z) = \frac{\fstar{}}{\dutycycle{}\,\hubbletime{}(z)}\left(\frac{\omegab{}}{\omegam{}}m \right)\,\msolar{}\,\yr{}^{-1},
\end{equation}
where $(\omegab/\omegam\,m)$ is the mass of baryons in the galaxy and $\dutycycle{}\,\hubbletime{}(z)$ is the total time over which the galaxy was forming stars.  The star formation rate is used to obtain the ionizing photon luminosity\footnote{Our model is insensitive to the exact relationship between \ionlum{} to \sfr{} as any changes to the assumed evolutionary synthesis model are absorbed into the free parameters \fstar{} and \fescape{} during the luminosity function fitting.  Thus the high-redshift, low-abundance ($Z=0.05\,\Zsolar{}$), model for \ionlum{} in \citet{Schaerer:2003p3229} produces the same results as the solar abundance model used in \citet{Kennicutt:1998p4954}, which for simplicity is used for the remainder of this paper.} (\ionlum{}) following \citet{Kennicutt:1998p4954}:
\begin{equation}
  \ionlum{} = 9.26 \times 10^{52} \sfr{}\,\second^{-1},
\end{equation}
which assumes a Salpeter IMF with stellar masses ranging $0.1-100\,\msolar{}$.  Assuming that two out of three ionizing photons that do not escape the galaxy are converted to \lya{} through case-B recombination \citep{Osterbrock:1989}, the final equation for the \lya{} luminosity is
\begin{equation}
  \label{eqn:lya_lum}
  \lyalum{} = 0.68\,\planck{}\, \lyafreq{}\left(1-\fescape{}\right)\ionlum{}\,\ergs{}\,\second{}^{-1},
\end{equation} 
where \planck{} is Planck's constant, \lyafreq{} is the frequency of a \lya{} photon, and \fescape{} is the fraction of ionizing photons that escape the galaxy without being absorbed.  Combining these equations gives us an expression for \lyalum{} that depends on the total halo mass and redshift, and is proportional to the model parameters \dutycycle{}, \fstar{}, and \fescape{}, which we assume to be mass independent.

\subsection[\texorpdfstring{\lya{}}{Ly-alpha} Transmission in the \texorpdfstring{\igm{}}{IGM}]{\texorpdfstring{$\bmath{\lya{}}$}{Ly-alpha} Transmission in the \texorpdfstring{\igm{}}{IGM}}
\label{sec:igm_transmission}
Observations of \lya{}-emitting galaxies are subject to resonant absorption of \lya{} flux from \metalline{H}{i} atoms in the \igm{}. \citet{Gunn:1965p1821} showed that for high-redshift \lya{} sources, even a modest neutral fraction of $10^{-5}$ can significantly reduce the number of transmitted \lya{} photons along our line of sight, as the photons blueward of \lyawl{} redshift through \lya{} resonance.  Therefore the transmission of \lya{} photons through the \igm{} is a critical component of a model for the \lya{} luminosity function.

To determine the impact that the \igm{} has on the transmission of \lya{} photons we make extensive use of the model outlined in \S3 of \citetalias{Dijkstra:2007p1}.  The \igm{} is modeled following the halo infall calculations of \citet{Barkana:2004p1709} and interpretation by \citetalias{Dijkstra:2007p1}, who provide a calculation of the density profile (\sub{\rho}{\igm{}}) and velocity field (\sub{v}{\igm{}}) of the \igm{} as a function of halo mass and distance from the halo [compare with \citealt{Dijkstra:2007p1} equation (4)]:
\begin{equation}
  \label{eqn:igm_properties}
  \begin{split}
    \sub{\rho}{\igm{}}(r,\sub{r}{vir})&=\begin{cases}
      20\,\bar{\rho}\,(r/\sub{r}{vir})^{-1} &r < 10\,\sub{r}{vir},\\
      \bar{\rho}\left[1 + \exp\left(2 - r/5\,\sub{r}{vir}\right)\right] &r \geq 10\,\sub{r}{vir},
    \end{cases} \\
    \sub{v}{\igm{}}(r,\sub{r}{vir}) &=\begin{cases}
       \left(\frac{r-\sub{r}{vir}}{9\,\sub{r}{vir}}\right)\left[H(z)\,r\right]- \sub{v}{circ}  &r < 10\,\sub{r}{vir},\\
       H(z)\,r &r \geq 10\,\sub{r}{vir},
    \end{cases}
  \end{split}
\end{equation}
where $r$ is the distance from the halo, \sub{r}{vir} and \sub{v}{circ} are the virial radius and circular velocity of the halo, and $H(z)$ is the Hubble parameter at redshift $z$.

The density profile is combined with the assumed photoionizing flux from the galaxy and the external UV background (\uvbg{}) to get a distance dependent neutral fraction (\nfrac{}).  These values are combined to give the total opacity of photons as a function of their wavelength $\lambda$ [compare with \citealt{Dijkstra:2007p1} equation (8)]:
\begin{equation}
  \tau(\lambda)=\int_{\sub{r}{vir}}^{\infty}\rmn{d}r\,\sub{\rho}{\igm{}}\left(r\right)\,\nfrac{}\left(r\right)\,\sub{\sigma}{\lya{}}\left[\lambda, \sub{v}{\igm{}}\left(r\right)\right],
\end{equation}
where \sub{\sigma}{\lya{}} is the absorption cross-section for the rotationally broadened \lya{} line, evaluated at wavelength $\lambda$, and blueshifted by the velocity field \sub{v}{\igm{}}.  This opacity is then convolved with an assumed \igm{} density fluctuation distribution \citep{MiraldaEscude:2000p2867} to account for any clumpy overdensities along the line of sight.  The resulting function $\wltrans{}(\lambda)$ gives the fraction of photons emitted at wavelength $\lambda$ that are transmitted through the \igm{} without being scattered out of the line of sight.

The quantity of interest is the total fraction of the \lya{} line transmitted through the \igm{}:
\begin{equation}
  \igmtrans{} = \frac{\int \rmn{d}\lambda\,\wltrans{}(\lambda)\,J(\lambda)}{\int \rmn{d}\lambda\,J(\lambda)},
\end{equation}
where $J(\lambda)$ is the flux of the galaxy as a function of wavelength. A galaxy with an intrinsic luminosity above the detection limit can be pushed below detectability for a sufficiently small value of \igmtrans{}.

\begin{figure}
  \centering 
  \includegraphics[width=0.95\columnwidth]{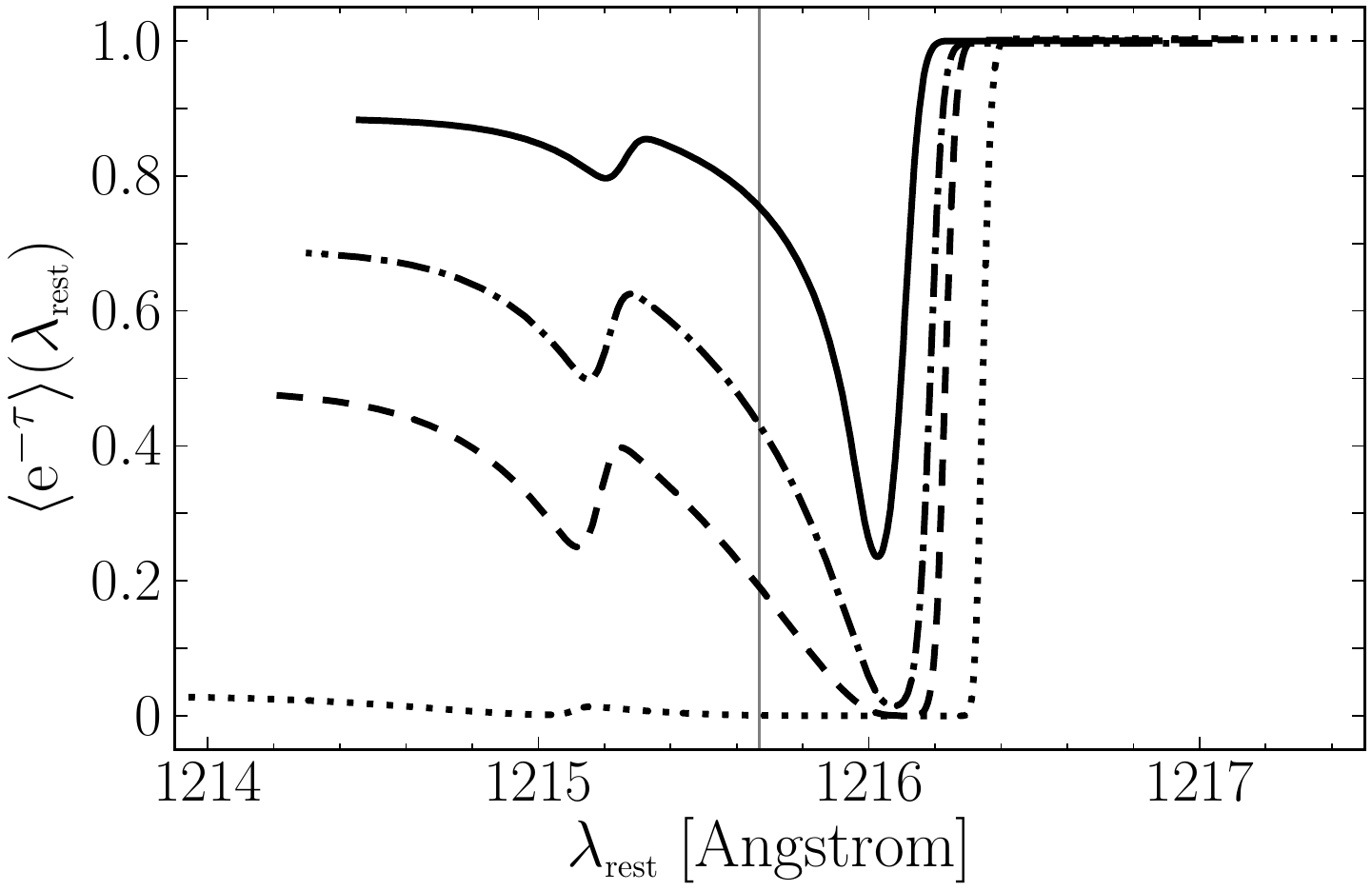} 
  \caption{The fraction of doppler-shifted \lya{} photons transmitted through the \igm{} from a galaxy with total mass $10^{11}\,\msolar{}$ as a function of wavelength for field conditions at redshifts (from top to bottom) $z=2.2$, $3.1$, $3.7$, and $5.7$.  The integrated transmission across the broadened \lya{} for each redshift is (from top to bottom) $\igmtrans{}=0.74$, $0.48$, $0.31$, $0.13$.  For reference the vertical line is the intrinsic \lyawl{}.}
  \label{fig:redshift_transmission}
\end{figure}

\figref{fig:redshift_transmission} shows $\wltrans{}(\lambda)$ across the width of the rest frame \lya{} line for a galaxy with mass $10^{11}\msolar{}$ in mean conditions at four redshifts.  The assumed shape of the broadened \lya{} line is Gaussian with its width set by the assumed dark matter circular velocity following \citetalias{Dijkstra:2007p1}\footnote{\citet{Dijkstra:2010p5735} show that galactic outflows of \metalline{H}{i} modify the \lya{} spectral line shape decreasing the impact of the \igm{} on transmission, implying that our simple Gaussian model is a conservative calculation of \igmtrans{}.}.  The four curves correspond to the redshifts (top to bottom) $z=2.2$, $3.1$, $3.7$, and $5.7$, with integrated transmission fractions (top to bottom) $\igmtrans{}=0.74$, $0.48$, $0.31$, and $0.13$.  For reference the vertical line is \lyawl{}.  The sharp drop in transmission around $1216\,\angstrom{}$ is caused by the blueshifted \lya{} resonance as seen by photons escaping the galaxy through infalling hydrogen gas from the \igm{}.  Between $1215$ and $1216\,\angstrom{}$ the galaxy's internal ionization and the ionizing background decrease the opacity of the nearby \igm{} to \lya{} photons causing the blueward rollup in $\wltrans{}(\lambda)$, and as the ionizing background and UV mean free path increases with decreasing redshift this effect becomes more pronounced.  Note that the total width of the broadened \lya{} line changes as the circular velocity of the galaxy changes with redshift, resulting in different wavelength spans.

The shape of the transmission curve changes with redshift as the model components evolve, with the biggest contributions coming from the decreasing mean hydrogen density which evolves from $5.8\times10^{-5}\,\centimeter^{-3}$ at $z=5.7$ to $6.1\times10^{-6}\,\centimeter^{-3}$ at $z=2.2$, and the external UV background which evolves from $0.34\times10^{-12}\,\second^{-1}$ at $z=5.7$ to $1.20\times10^{-12}\,\second^{-1}$ at $z=2.2$ \citep{Bolton:2007p4060}.  The decrease in \igm{} density and increase in \uvbg{} allows for the \igm{} transmission fraction to increase by a factor of six between redshifts $5.7$ and $2.2$.  This illustrates the sensitivity of the \lya{} transmission to these two environmental factors, and motivates the idea that the quasar environment may be significantly reducing the transmission of nearby \lya{} emitters.

\begin{figure}
  \centering 
  \includegraphics[width=0.95\columnwidth]{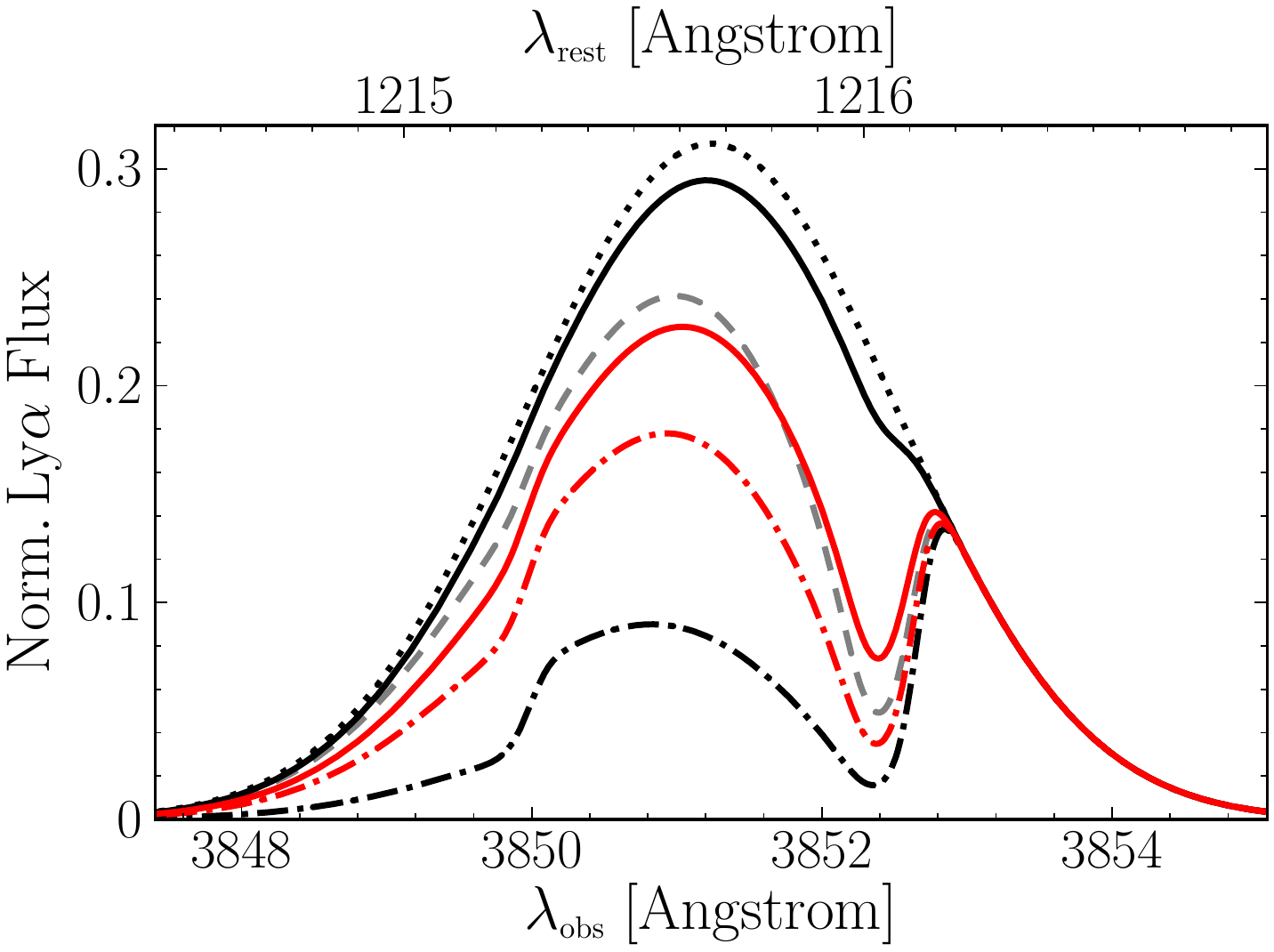} 
  \caption{\lya{} line shapes for a $10^{11}\,\msolar{}$ galaxy at $z=2.2$ in various environments.  The dotted curve is the original \lya{} line shape and the grey dashed curve is the transmitted line shape assuming a mean \igm{} density and UV background.  The solid \blackbold{} (\redbold{}) curves show the impact of an increased \igm{} density and UV flux from the nearby quasar with a host halo size of $2\times10^{13}\,\msolar{}$ ($10^{12}\,\msolar{}$).  The dash-dotted curves include the increased \igm{} density but without the quasar's contribution to the incident UV flux.}
  \label{fig:galaxy_transmission}
\end{figure}

\figref{fig:galaxy_transmission} shows $\wltrans{}(\lambda)$ convolved with an assumed \lya{} line profile to simulate the continuum subtracted spectrum for a galaxy with total mass $10^{11}\,\msolar{}$ at $z=2.2$.  The dotted curve is the original \lya{} line shape, and the grey dashed curve is the transmitted line shape assuming a mean \igm{} density and UV background. The sharp feature around $3852\,\angstrom{}$ is the blueshifted \lya{} resonance as seen by photons escaping the galaxy through infalling hydrogen gas from the \igm{}.  In the absence of a strong UV flux the entire line blueward of this shifted resonance is scattered out of the line of sight and the transmission is lower than the $0.5$ predicted by \citet{Gunn:1965p1821}.  With a large enough UV flux from internal (the galaxy) and external (UV background) sources, the neutral fraction of the \igm{} in the vicinity of the galaxy is lowered considerably, and as a result, the transmission of the galaxy's intrinsic \lya{} luminosity is increased.

\subsection[Galactic UV Luminosity]{Galactic UV Luminosity}
\label{sec:uv_lum}
In addition to \lya{} luminosity our model also calculates a UV magnitude for each galaxy.  This allows us to use an additional dataset to help constrain \dutycycle{}, \fstar{}, and \fescape{} at the cost of an additional fitting parameter \fdust{} to account for UV-specific\footnote{We do not explicitly account for dust extinction in our treatment of \lyalum{} as any pre-\igm{} effects of dust on \lyalum{} is absorbed into \fescape{} while fitting the luminosity functions.} dust extinction.  We assume this to be luminosity independent and fit this as an additional free parameter in our model \citep{Dayal:2011p6072}. Our fitted values for \fdust{} match those found by \citet{Bouwens:2009p5032}.

We use the relationship between star formation rate and rest-frame UV continuum luminosity (\uvcontlum{}) as calculated by \citet{Kennicutt:1998p4954}:
\begin{equation}
  \uvcontlum{} = 7.1 \times 10^{27} \fdust{} \sfr{}\,\ergs{}\,\second{}^{-1}\,\hertz{}^{-1}.
\end{equation}
Combining this expression with \eqnref{eqn:star_formation_rate} gives us the relationship between UV luminosity and the free model parameters.  To compare with the UV luminosity function in \citetalias{Ouchi:2008p3716}, this is converted to an absolute AB magnitude:
\begin{equation}
  \label{eqn:uv_mag}
  \sub{\rmn{M}}{AB} = -2.5\,\logten{}\left(\uvcontlum{}\right) + 51.6.
\end{equation}

\subsection[Fitting UV and \texorpdfstring{\lya{}}{Ly-alpha} Luminosity Functions]{Fitting UV and \texorpdfstring{$\bmath{\lya{}}$}{Ly-alpha} Luminosity Functions}
\label{sec:luminosity_functions}
To constrain our \lya{} model parameters \dutycycle{}, \fstar{}, and \fescape{} we fit the  $z=3.1$ \lya{} and UV differential luminosity functions presented in \citetalias{Ouchi:2008p3716} (their figures 16 and 22 respectively) using the luminosity model presented above.  For each luminosity bin in the observed \lya{} and UV luminosity functions we determine the mass required to generate that luminosity by inverting \eqnref{eqn:lya_lum} and \eqnref{eqn:uv_mag} respectively.  This mass is used to obtain the halo number density for that luminosity bin using the extended Press-Schechter mass function, $\rmn{d}n/\rmn{d}m$ \citep{Sheth:2001p1889}.  The halo number density is then converted to a galaxy number density by assuming only a fraction \dutycycle{} of the halos are occupied by star forming galaxies in the observed epoch and therefore emitting a detectable \lyalum{} and \uvcontlum{}.  This yields the following equations for the differential \lya{} and UV luminosity functions:
\begin{align}
  \label{eqn:lum_funcs}
  \sub{n}{\lya{}}(\igmtrans{}\times\lyalum{}) &= \dutycycle{}\,\frac{\rmn{d}n}{\rmn{d}m}\frac{\rmn{d}m}{\rmn{d}\logten{}L}, \\
  \sub{n}{\rmn{UV}}(\sub{M}{UV}) &= \dutycycle{}\,\frac{\rmn{d}n}{\rmn{d}m}\frac{\rmn{d}m}{\rmn{d}\sub{M}{UV}},
\end{align}
where $\sub{n}{\lya{}}$ is the number density of \lya{} galaxies with luminosities between $\logten{}L$ and $\logten{}L + \rmn{d}\logten{}L$, and $\sub{n}{\rmn{UV}}$ is the number density of UV galaxies with AB magnitudes between $\sub{M}{UV}$ and $\sub{M}{UV}+\rmn{d}\sub{M}{UV}$. 

Both the \lya{} and UV luminosity functions are fit simultaneously by comparing to the data points in \citetalias{Ouchi:2008p3716} and computing a $\chi^{2}$ for both \lya{} and UV.  The combined value $\sub{\chi}{tot}^{2}=\sub{\chi}{\lya{}}^{2} + \sub{\chi}{UV}^{2}$ is minimized by varying \fstar{}, \fescape{} (\lya{}), \fdust{} (UV) for fixed values of \dutycycle{}.  The parameter \dutycycle{} is highly degenerate in our model and so the resultant best fits for a range of values are presented in \tblref{tbl:model_best_fit}\footnote{Our model assumes long-term, continuous, galactic star formation for simplicity. Recent work by \citet{Sharp:2010p5603} showed that star formation is impulsive on timescales less than $<10^{7}$ years and that for active galactic nuclei the timescales are much longer ($>10^{7}\,\yr{}$) corresponding to a $>3\,\cmpc{}$ light crossing time between cycles.  This indicates that more sophisticated simulations will need a galactic duty cycle at least several times $10^{7}\,\yr{}$ to be consistent with observation and that our modeling of a quasar's ionizing field as a continuous event is justified in our simulated volume size.}.

\begin{table}
  \caption{Best fit parameters for model fitting \lya{} and UV luminosity functions in \citetalias{Ouchi:2008p3716}.}
  \label{tbl:model_best_fit}
  \begin{tabular}{@{}cccccc}
    \hline
    \dutycycle{} & \fstar{}$\scriptstyle\times10^{-2}$ 
                 & \fescape{}$\scriptstyle\times10^{-1}$ 
                 & \fdust{} 
                 & $\sub{\chi}{tot}^{2} / d{\,}^{\alpha}$ 
                 & $\chi^{2} \gtrsim \sub{\chi}{tot}^{2}{}^{\beta}$ \\
    \hline
    0.4  &  7.7  &  1.45  &  0.22  &  1.90  &   4\% \\
    0.6  &  8.4  &  1.15  &  0.24  &  1.32  &  21\% \\
    0.8  &  8.9  &  0.81  &  0.26  &  1.00  &  44\% \\
    1.0  &  9.3  &  0.56  &  0.28  &  0.82  &  61\% \\
    \hline
  \end{tabular}

  \medskip
  $^{\alpha}$ $d=10$ is the degrees of freedom in the fit.\\
  $^{\beta}$ Probability that a random $\chi^{2}$ will be greater than $\sub{\chi}{tot}^{2}$.
\end{table}

\begin{figure*}
  \centering 
  \includegraphics[width=\textwidth]{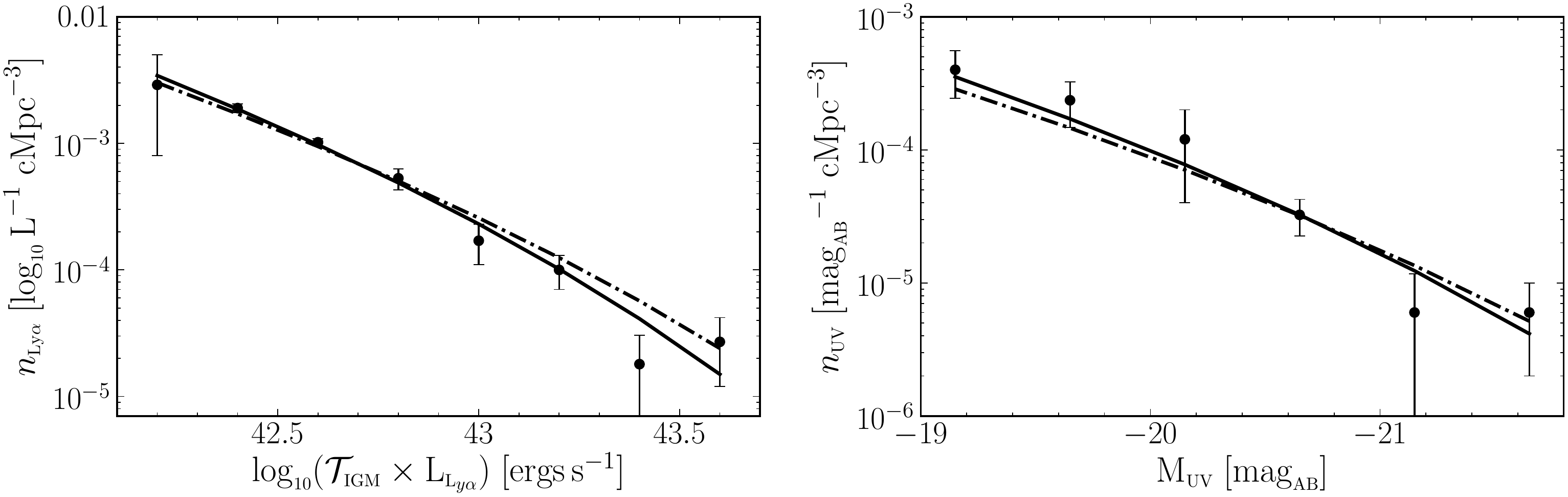} 
  \caption{\lya{} (left panel) and UV (right panel) fits to $z=3.1$ differential luminosity functions for two values of \dutycycle{}.  The points and error bars are from \citetalias{Ouchi:2008p3716}, figures 16 (left panel) and 22 (right panel).  The best and worst fit values of \dutycycle{} are plotted, $\dutycycle{} = 1.0$ (solid curves) and $\dutycycle{} = 0.4$ (dash-dotted curves), with a reduced $\sub{\chi}{tot}^{2}$ of $0.82$ and $1.90$ respectively. Intermediate values of \dutycycle{} produce fits that lie between these two curves.}
  \label{fig:luminosity_functions}
\end{figure*}

\figref{fig:luminosity_functions} shows our model fits to the $z=3.1$ \lya{} and UV luminosity functions presented in \citetalias{Ouchi:2008p3716}.  The fits correspond to the upper and lower bounds of $\dutycycle{}=0.4$ and $1.0$ that are in agreement with the data.  There are $14$ binned data points in total, $8$ for the \lya{} luminosity function and $6$ for the UV.  This is fit with the $4$ fitting parameters for a total of $10$ degrees of freedom.  We found that $\dutycycle{}=1.0$ yields the best fit for the luminosity functions, with a reduced $\sub{\chi}{tot}^{2}$ of $0.82$.  This corresponds to a probability of $61\%$ that a randomized set of parameters would produced a larger $\sub{\chi}{tot}^{2}$, and we use this value of \dutycycle{} for all calculations on the impact of a quasar environment in the following section. The quality of fit decreased monotonically with \dutycycle{} and we found that below a $0.4$ the fit was in significant disagreement ($<4\%$) with the data.  We will show in \secref{sec:comparison} the exact choice of \dutycycle{} does not change our main result.

As a further check we compared our calculated \lya{} equivalent widths (\equivwidth{}) to the mean observed at $z=3.1$ in \citetalias{Ouchi:2008p3716}.  From their spectroscopic and photometric samples \citetalias{Ouchi:2008p3716} calculated a mean \equivwidth{} of $102$ and $130\,\angstrom{}$ respectively.  Our best fit model has a mean \equivwidth{} of $120\,\angstrom{}$, in good agreement with these observations.  Moreover, because $\equivwidth{} \propto \igmtrans{}(1-\fescape{})/\fdust{}$ this is independent evidence of a robust fit of our parameters to observation.  Our range of $\fdust{}=0.22$ to $0.28$, also matches nicely with recent results in \citet{Blanc:2010p5766} who found this parameter to be between $0.20$ to $0.30$.

We note that all of the constraints on our parameters are obtained by fitting to the $z=3.1$ luminosity functions observed in \citetalias{Ouchi:2008p3716}, but are then applied to an environment at $z=2.2$.  Current \lya{} luminosity functions at $z<3$ still suffer from cosmic variance \citep[see][and references therein]{Blanc:2010p5766} and there exist no accompanying UV luminosity functions at these lower redshifts.  Therefore fitting to data from $z=3.1$ is the best that can be done to constrain the free parameters in the model at lower redshift.

It is therefore important to note that recent observations do show evolving \lae{} properties with redshift, in particular the \equivwidth{} distribution at $z=2.3$ \citep{Nilsson:2009p6059} is described by a steeper exponential function than at $z=3.1$ \citep{Gronwall:2007p3702} yielding a base \equivwidth{} at $z=2.3$ that is roughly two-thirds of the value found at $z=3.1$.  This suggests that \lya{} photons are more difficult to detect relative to the UV continuum from $z=2.3$ galaxies than those emitted from $z=3.1$.  Therefore, to match the smaller $z=2.3$ mean \equivwidth{} would require either an increase in our \fdust{} and/or \fescape{} parameters.  Without a matching set of \lya{} and UV luminosity functions at this lower redshift we can only conjecture on exactly how this would affect our free parameters.  Changing only \fescape{} to match this reduced \equivwidth{} would have the largest effect on our results.  We therefore ran our simulations with this lower \fescape{} and found that the detected number of \lae{} was lowered by $35\%$, but that this did not change our final results.  For the rest of the paper the fitted parameters in \tblref{tbl:model_best_fit} are used.

\subsection{Quasar Influenced Transmission}
\label{sec:quasar_transmission}
The additional component of our model beyond the work of \citetalias{Dijkstra:2007p1} concerns the environmental effects contributed by a nearby quasar,  including an enhanced \sub{\rho}{\igm{}} and \uvbg{}.  These have a significant impact on the \lya{} line shape, and are computed semi-analytically from the quasar's observed $B$-band apparent magnitude, \sub{m}{B}, and its redshift, \sub{z}{Q}.  To determine both the increased ionization rate and \igm{} density in the vicinity of the quasar we first calculate its $B$-band luminosity \sub{L}{B}.  Assuming that the quasar is shining at the Eddington limit we then find the black hole mass required to emit at the observed luminosity, and in turn calculate the expected host halo size using the conversion in \citet{Wyithe:2005p322}.  

To calculate the effective density of the \igm{} in the vicinity of a galaxy, we first write \eqnref{eqn:igm_properties} as a density excess relative to the mean density $\bar{\rho}$:
\begin{equation}
  \Delta\rho(r,\sub{r}{vir}) = \sub{\rho}{\igm{}}(r,\sub{r}{vir}) - \bar{\rho}.
\end{equation}
This density excess is then added to the underlying density contribution of the quasar to get the total effective density profile used in calculating \igmtrans{}.  For a galaxy at a distance \sub{r}{Q} from a quasar with viral radius \sub{r}{Q,vir}, the combined density of the local \igm{} a distance $r$ from a galaxy with virial radius \sub{r}{vir} is:
\begin{equation}
  \sub{\rho}{gal}(r,\sub{r}{vir}) = \sub{\rho}{\igm{}}(\sub{r}{Q},\sub{r}{Q,vir}) + \Delta\rho(r,\sub{r}{vir}).
\end{equation}
With this formulation, a galaxy just outside the virial radius of the quasar ($\sub{r}{Q}=\sub{r}{Q,vir}$) has an effective density that ranges from a maximum value of $39\,\bar{\rho}$ at the galaxy's virial radius ($r=\sub{r}{vir}$), to a minimum of $20\,\bar{\rho}$ at large distances from the galaxy ($r\gtrsim 30\,\sub{r}{vir}$).  For a galaxy well away from the quasar ($\sub{r}{Q}\gtrsim 30\,\sub{r}{Q,vir}$), the effective density ranges from $20\,\bar{\rho}$ at the galaxy's virial radius to the mean \igm{} density $\bar{\rho}$ at large distances.

To calculate the ionization rate of the quasar we follow the prescription of \citet{Schirber:2003p483} and determine the quasar's flux at the Lyman limit (\sub{J}{LL}) given its observed \sub{L}{B}.  This, along with the assumed UV-continuum slope of the quasar and the cross-section of hydrogen, is used to calculate the number of hydrogen ionizations per second (\sub{\Gamma}{Q}) as a function of quasar luminosity and distance from the quasar:
\begin{multline}
  \label{eqn:quasar_flux}
  \sub{\Gamma}{Q}(r,\sub{L}{B}) = \frac{12.0}{3+\sub{\alpha}{UV}} \left( \frac{\sub{J}{LL}(r, \sub{L}{B})}{\scriptstyle 10^{-21}\,\ergs{}\,\second{}^{-1}\centimeter{}^{-2}\,\hertz{}^{-1}\,\steradian{}^{-1}} \right)\\
  \times\exp\left(-r/\mfp{}\right)\,10^{-12}\,\second{}^{-1},
\end{multline}
where \sub{\alpha}{UV} is the UV-continuum slope of the quasar, and \mfp{} is the inferred mean free path of UV photons attenuating the ionizing flux \citep{FaucherGiguere:2008p4083}.  We generate the ionization field for each galaxy using a two component model consisting of the distance dependent quasar flux in \eqnref{eqn:quasar_flux} and a constant ionizing background.\footnote{This assumption was compared with a numerical model that treated each galaxy discretely within the quasar's zone of influence ($\sub{r}{Q}\lesssim 30\,\sub{r}{Q,vir}$) and used a smoothed ionizing luminosity from $30\,\sub{r}{Q,vir}$ to $10\,\mfp{}$.  The sum of the galaxy contribution and the smoothed background was flat across the entire volume, with the clustering of galaxies pushing the flux up by $\sim2\%$ at the center of the simulation volume. This justifies the simple two component model of the effective ionizing flux.}

As an example, \figref{fig:galaxy_transmission} shows the environmental effects on transmission for a $10^{11}\,\msolar{}$ galaxy $500\,\pkpc{}$ from a nearby quasar.  The black curves assume a central quasar matching the properties observed in \citetalias{Francis:2004p2989}, with an apparent $B$-band magnitude of $17.6$, a corresponding $B$-band luminosity of $4.4\times10^{13}\,\lbsolar{}$, and inferred black hole and host halo masses of $7.6\times10^{9}\,\msolar{}$ and  $2\times10^{13}\,\msolar{}$ respectively.  The UV flux generated by the quasar in the vicinity of the galaxy very nearly removes the \igm{}'s effect on the \lya{} line, yielding a transmission of $0.94$, seen in the solid black curve.  The black dash-dotted curve shows how the enhanced density of the \igm{} decreases the transmission when the UV flux is not enhanced by the quasar, resulting in a final transmission of $0.33$.  This is compared to the dashed curve showing the \lya{} line assuming a mean \igm{} and background UV flux, with a total transmission of $0.74$.

For comparison, the red curves show the same galaxy in the vicinity of a much smaller and less luminous quasar, with an apparent $B$-band magnitude of $23.0$, a corresponding $B$-band luminosity of $3\times10^{11}\,\lbsolar{}$, and inferred black hole and host halo masses of $5.3\times10^{7}\,\msolar{}$ and  $10^{12}\,\msolar{}$ respectively.  For this smaller quasar the UV flux is only sufficient to compensate for the increased density of the \igm{}, as seen in the solid red curve which roughly matches the dashed background curve with a total transmission of $0.71$.  Because the quasar has a smaller host halo the density boost to the \igm{} is not as strong at the same radius, which allows for a transmission of $0.56$ in the absence of the quasar's UV flux seen in the red dash-dotted curve.

\begin{figure}
  \centering 
  \includegraphics[width=0.95\columnwidth]{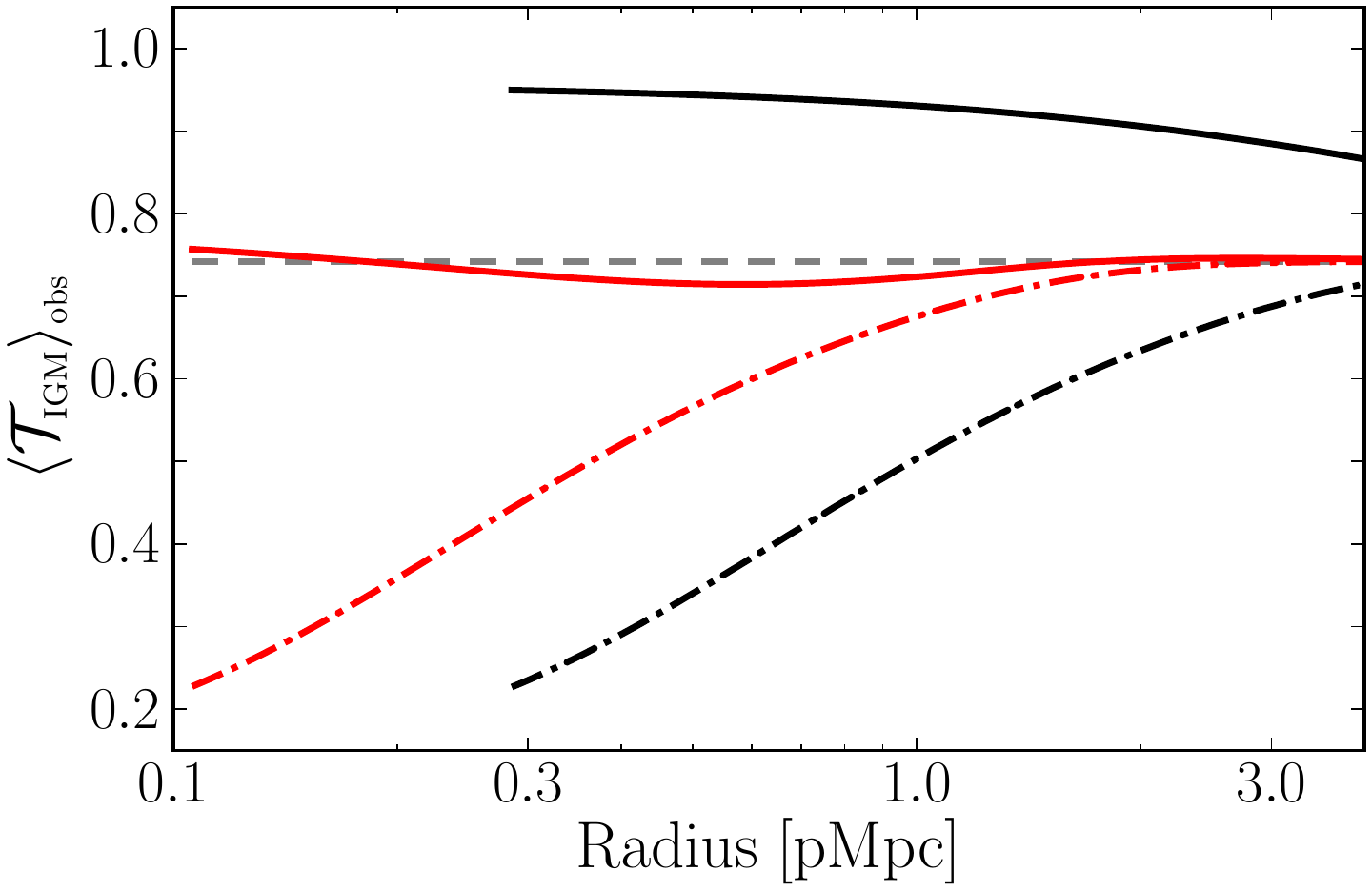} 
  \caption{Averaged \lya{} transmission as a function of radius from a central quasar with host halo size (\blackbold{} curves) $2\times10^{13}\,\msolar{}$ and (\redbold{} curves) $10^{12}\,\msolar{}$. The grey dashed curve is the average transmission assuming a mean \igm{} density and UV flux.  The solid curves include the density and UV flux from the nearby quasar and the dash-dotted curves include the increased density but with no UV boost.}
  \label{fig:quasar_transmission}
\end{figure}

The exact shape of the radiation fields emitted from a quasar is an open and thorny question.  If the emissions are powerful and tightly collimated the ionization is still likely to be diffused through the volume in some way, either by scattering off of dust grains or electrons via Thompson scattering \citep{BlandHawthorn:1991p6365, Sokolowski:1991p6361}.  Moreover, if the beam moves with time then a time-averaged UV field must be considered.  The impact of these details can be averaged over by analyzing large surveys containing observations like that in \citetalias{Francis:2004p2989}.  This allows for the limiting cases to be described by a dilution factor \fdilution{} representing the fractional solid angle that the radiation field emitted by the quasar permeates, where $\fdilution{}=1$ corresponds to a perfectly isotropic quasar and $\fdilution{}=0$ to a quasar emitting no ionizing radiation.  For the purposes of this paper, in which we analyze the results of a single observation, we will only look at these two extremes.

\figref{fig:quasar_transmission} shows the \lya{} transmission averaged over observable galaxies, assuming the \citetalias{Francis:2004p2989} detection limit, as a function of distance from the quasar at $z=2.2$.  As in \figref{fig:galaxy_transmission}, the black and red lines represent a central quasar with host halo mass $2\times10^{13}\,\msolar{}$ and $10^{12}\,\msolar{}$ respectively, the solid and dash-dotted lines show the effects of the quasar with ($\fdilution{}=1$) and without ($\fdilution{}=0$) the enhanced ionizing flux. The dashed grey line is the average transmission for a mean \igm{}.  For each quasar the virial radius was used as the minimum radial distance in the model to avoid confusion in halo occupation; the larger and smaller quasars have a minimum radius of $280\,\pkpc{}$ and $100\,\pkpc{}$ respectively.

For the smaller quasar we see that \sub{v}{\igm{}} and \sub{\Gamma}{Q} compensate for one another, leading the a radial transmission curve that is nearly equal to that of mean \igm{}.  The larger quasar has a much larger UV flux and is able to ionize the denser \igm{} efficiently, leading to a transmission roughly $20\%$ higher than the mean.  Assuming that a quasar's ionizing flux is isotropic, this implies that the transmission in the vicinity of a quasar at $z=2.2$ is comparable to the mean \igm{} for modest sized quasars and is higher for the very luminous.  Therefore if \igmtrans{} is the primary factor influencing detection of \lya{} emission from galaxies near to a luminous quasar, we then would expect to see an {\em increase} in the number of observable \lya{} galaxies in the vicinity of the quasar observed by \citetalias{Francis:2004p2989}.

\section[Comparison With Observations Around \texorpdfstring{\jossquasar{}}{PKS 0424-131}]{Comparison With Observations Around \texorpdfstring{\jossquasar{}}{PKS 0424-131}}
\label{sec:comparison}

Having described the model, we are now in a position to interpret the observations of \jossquasar{} \citepalias{Francis:2004p2989}.  We construct a physical volume around \jossquasar{} matching the observed volume, and fill it with galaxies following our semi-analytic prescription.  We do this by first building a series of concentric spherical shells centered on our quasar's modeled halo, distributed evenly in log-space.  The shell radii range from the quasar halo's virial radius out to a radius that encloses the observed volume.  These shells are then filled with dark matter halos over a range of masses as given by the \citet{Press:1974p1874} mass function \citep[as modified by][]{Sheth:2001p1889}, enhanced by the dark matter two-point correlation function with the linear bias of \citet{MO:1996p2403}. We set the minimum halo mass of the model to produce an intrinsic luminosity \lyalum{} at the survey detection limit.  These halos are populated with galaxies as in \secref{sec:luminosity_functions}, by assuming all the halos are populated but that only a fraction \dutycycle{} of these galaxies are star forming (and thus detectable) in the observed epoch.  We use the best fit parameters based on our analysis of the field luminosity functions.

Thus the average number of halos $\sub{\mathcal{N}}{gal}(m,r)$, with mass in the range $m\pm\rmn{d}m/2$, with a separation of $r\pm\rmn{d}r/2$ from a central quasar of mass $M$ at redshift $z$ is: 
\begin{multline}
  \label{eqn:average_galaxies}
  \sub{\mathcal{N}}{gal}(m,r) = \dutycycle{} (4\pi r^{2}\rmn{d}r) \\
  \times \frac{\rmn{d}n(m, z)}{\rmn{d}m}\left[1 + \xi(r, m, M, z)\right],
\end{multline}
where $\rmn{d}n/\rmn{d}m$ is the halo mass function, and $\xi$ is the linearly biased two-point correlation function. This is the prescription found in $\S2.1.1$ of \citet{Dijkstra:2008p575} [\eqnref{eqn:average_galaxies} is a modification of their equation (1)], which describes the clustering of galaxies around a central dark matter halo but assumes a non-linear treatment of the separation bias.  As we are probing much larger separations a linear treatment is sufficient for our purposes.

\citetalias{Francis:2004p2989} surveyed a field of view measuring $4.2\,\pmpc{}$ by $2.0\,\pmpc{}$ over a redshift range of $\Delta z=0.015$ centered on the $z=2.168$ luminous quasar \jossquasar{}.  This range corresponds to a physical depth of $6.5\,\pmpc{}$ at this redshift and an effective depth of $8.5\,\pmpc{}$ when accounting for peculiar velocity induced redshift distortion \citep{Kaiser:1987p3306}. This yields an effective survey volume of $70\,\pmpc{}^{3}$.

Their detection limit for \lya{} emitting galaxies within this volume was $9.6\times10^{-18}\,\ergs{}\,\centimeter{}^{-2}\,\second{}^{-2}$, corresponding to an ideal ($\igmtrans{}=1$) minimum mass between $3.8\times10^{10}\,\msolar{}$ ($\dutycycle{}=0.4$) and $7.1\times10^{10}\,\msolar{}$ ($\dutycycle{}=1.0$).  In our simulation we consider galaxies with host halos ranging from this minimum detectable mass to a maximum of $10^{15}\,\msolar{}$ for completeness, and use $500$ mass bins distributed equally in log-space to span this range.  Using the inferred mass of $2\times10^{13}\,\msolar{}$ for \jossquasar{} described in \secref{sec:quasar_transmission} we calculated a virial radius of $280\,\pkpc{}$ which we used for the innermost spherical shell.  The outermost shell was given a radius $4.84\,\pmpc{}$, fully enclosing the rectangular observed volume.  We use $100$ radial shells to span this range of radii, distributed equally in log-space.

For each of these radial shells we calculate the quasar's UV flux from \eqnref{eqn:quasar_flux} and combine it with the external UV background to give an effective ionizing rate for the shell, and determine the quasar-enhanced \igm{} density and \igm{} neutral fraction detailed in \secref{sec:quasar_transmission}.  For each of the mass bins within each shell we determine the transmission given the quasar-influenced conditions and determine which masses have a detectable luminosity post-transmission.  For the masses that are above this threshold we sum the average number of galaxies expected at this separation given by \eqnref{eqn:average_galaxies}.  The total number of galaxies for each radial bin are added to give a final value for expected average enclosed galaxies within the survey volume.

Because our model is generated from spherical shells and the effective observed volume around \jossquasar{} is rectangular with dimensional ratios of roughly $4$:$2$:$1$, we determine for each shell the fractional volume that is located outside the observed rectangular volume and reduce that shell's contribution to the enclosed galaxy count accordingly.  This properly treats the radially dependent clustering and UV flux in both the long and short axes of the observed volume.

\begin{figure}
  \centering 
  \includegraphics[width=0.95\columnwidth]{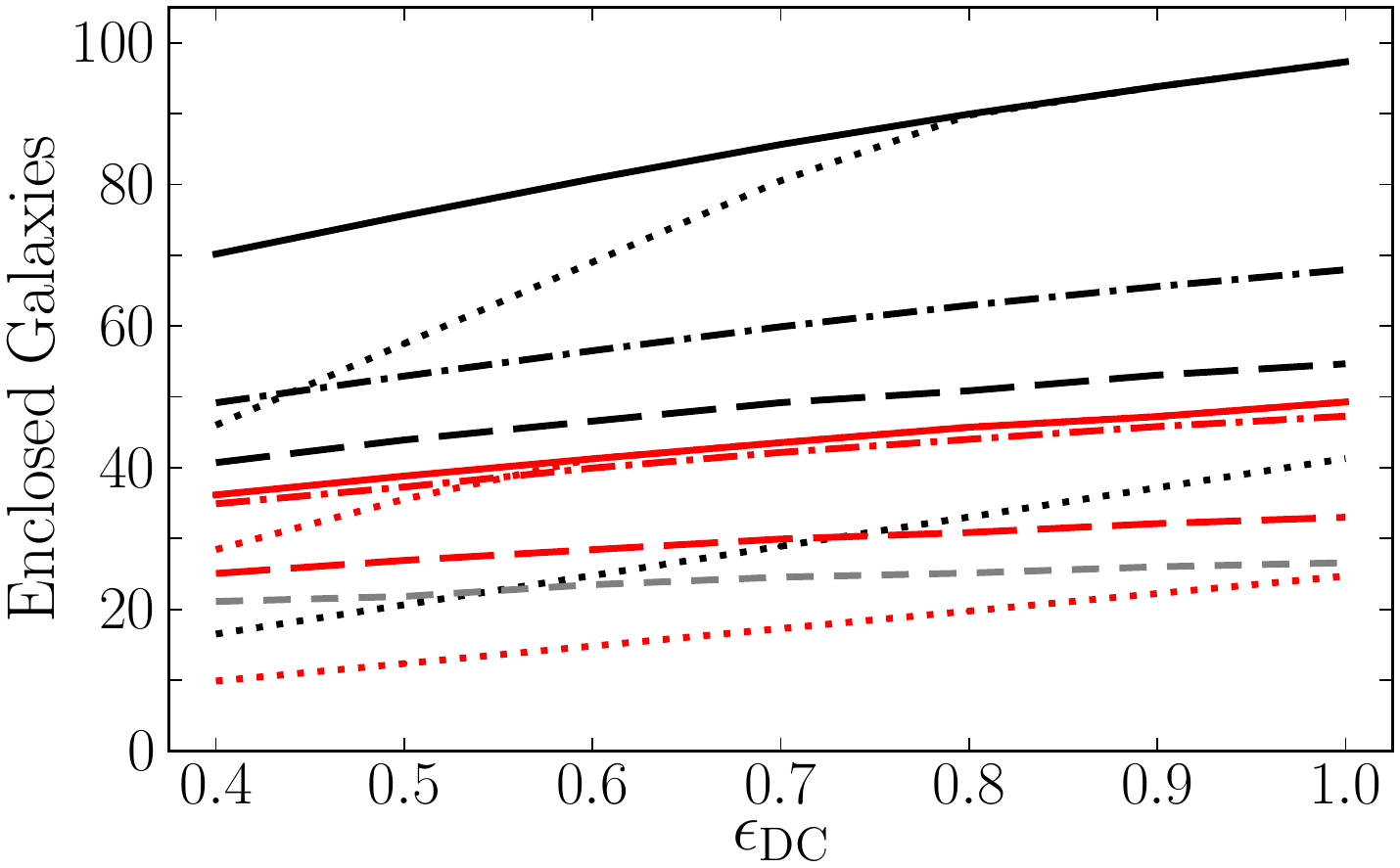} 
  \caption{Predicted number of galaxies contained within the Francis--Bland-Hawthorn volume as a function of modeled duty cycle for a quasar with host halo size (\blackbold{} curves) $2\times10^{13}\,\msolar{}$ and (\redbold{} curves) $10^{12}\,\msolar{}$.  The grey dashed curve is the number of galaxies expected in the field.  The long-dashed lines show the enclosed count assuming a constant $\igmtrans{}=0.5$. The solid (dash-dotted) curves include clustering and transmission effects including (excluding) the boosted UV flux from the quasar.  The upper and lower dotted curves for each quasar show the enclosed count excluding galaxies with $\virialtemp{} < 5\times10^{5}\,\kelvin{}$ ($\lesssim6.6\times10^{10}\,\msolar{}$) and $\virialtemp{} < 10^{6}\,\kelvin{}$ ($\lesssim1.9\times10^{11}\,\msolar{}$) respectively.}
  \label{fig:enclosed_galaxies}
\end{figure}

\figref{fig:enclosed_galaxies} shows the results for this process over the allowed range of \dutycycle{}.  The solid black curve shows that the average number of galaxies that should be visible within the volume surveyed by \citetalias{Francis:2004p2989} given the fiducial model including the quasar is between $70$ and $97$.  Even if the ionizing flux from the quasar itself is omitted, as seen in the dash-dotted black curve, the average number of visible galaxies is still between $49$ and $67$.  This can be contrasted with the grey dashed line which gives the expected number of galaxies assuming a mean density and \igm{} with no quasar, which ranges between $21$ and $26$, and the long-dashed black line which gives the galaxy count assuming a constant $\igmtrans{}=0.5$, ranging between $40$ and $54$.  The clustering of galaxies around the quasar, combined with the higher than background transmission rates seen in \figref{fig:quasar_transmission}, pushes the expected number of galaxies up well above background levels, in contrast with observation.

We have also repeated the above calculation for the same observable volume and redshift but for a much less luminous central quasar.  The mass of the dark matter halo hosting the quasar was set to $10^{12}\,\msolar{}$ and the innermost spherical shell's radius to this quasar's lower virial radius of $100\,\pkpc{}$, keeping the rest of the conditions the same.  In this case the lower \igm{} density and lower UV flux conspire to give close to the same number of galaxies for conditions when the quasar UV flux is considered and when it is neglected. The solid and dash-dotted red lines show the expected galaxy count in these cases, which varies between $34$ and $49$ over the allowed range in \dutycycle{}.  The long-dashed red line gives the galaxy count assuming a constant $\igmtrans{}=0.5$, ranging between $25$ and $33$.

Thus our model predicts a number of galaxies that is far in excess of the null-detection in \citetalias{Francis:2004p2989}.  This points to either an additional suppression of the \lya{} signal that has not been modeled, or else to some mechanism for the suppression of star formation.   To investigate this possibility we introduce a simple cutoff in mass below which galaxies are unobservable.  The dotted curves in \figref{fig:enclosed_galaxies} show the results from these imposed mass cuts, which correspond to $\virialtemp{} \leq 5.5\times10^{5}\,\kelvin{}$ [$\lesssim7.6\times10^{10}\,\msolar{}$] and $\virialtemp{} \leq 8.4\times10^{5}\,\kelvin{}$ [$\lesssim1.4\times10^{11}\,\msolar{}$]. In our model these halos yield the \lya{} flux limit with a constant $\igmtrans{}=0.5$ for $\dutycycle{}=0.4$ and $\dutycycle{}=1.0$ respectively.  The top black (red) dotted curve shows the enclosed count excluding galaxies with $\virialtemp{} \le 5\times10^{5}\,\kelvin{}$ [$\lesssim6.6\times10^{10}\,\msolar{}$] and the bottom black (red) dotted curve shows the enclosed count excluding galaxies with $\virialtemp{} \leq 10^{6}\,\kelvin{}$ [$\lesssim1.9\times10^{11}\,\msolar{}$].  The range of enclosed galaxies for the $\virialtemp{} \le 5\times10^{5}\,\kelvin{}$ cut runs from $46$ to $97$ for the larger quasar (upper black dotted curve) and from $28$ to $49$ for the smaller (upper red dotted curve).  For the $\virialtemp{} \leq 10^{6}\,\kelvin{}$ cut the enclosed count runs from $16$ to $41$ for the larger quasar (lower black dotted curve) and from $9$ to $24$ for the smaller quasar (lower red dotted curve). 

The mean number of galaxies required for the \citetalias{Francis:2004p2989} null detection around \jossquasar{} to be consistent with our model at the $68\%$ level is $\sim2.3$, and at the $90\%$ level is $\sim1.1$.  Thus, for $\dutycycle{}=0.4$ ($\dutycycle{}=1.0$) our model must exclude masses below $1.2\times10^{12}\,\msolar{}$ ($2.5\times10^{12}\,\msolar{}$) in order to bring the mean number of galaxies down to a value consistent with observations at the $68\%$ level, and must exclude masses below $2.1\times10^{12}\,\msolar{}$ ($4.2\times10^{12}\,\msolar{}$) to be consistent at the $90\%$ level.  This means that the most conservative virial temperature consistent with observation is $\virialtemp{} \gtrsim 3.4\times\,10^{6}\,\kelvin{}$ ($\sim1.2\times10^{12}\,\msolar{}$).  This result may imply considerable radiative suppression of galaxy formation by the nearby quasar.

\subsection{Additional Comparisons}
Several other studies have investigated the population of \lya{} emission near luminous quasars.  \citet{Cantalupo:2007p5794} reported a detection of $13$ \lya{} sources clustered around a quasar at $z=3.1$ in a $\sim25\,\pmpc{}^{3}$ volume (sparsely sampled from a larger $\sim200\,\pmpc{}^{3}$ volume), two of which they suggested were hydrogen clouds fluoresced by the quasar's ionizing radiation.  Our model, accounting for the sparse sampling, predicts approximately $20$ galaxies should be detected in this volume.  The possible externally fluoresced sources suggest there is enough ionizing flux around these high-redshift luminous to strongly impact neighboring \metalline{H}{i}.

\citet{Kashikawa:2007p5789} conducted a deep wide field narrowband survey for Lyman break galaxies and \lae{}s around QSO SDSS J0211-0009 at $z=4.9$.  They surveyed $\sim830\,\pmpc{}^{3}$ in the vicinity of the quasar and detected $221$ \lae{}s.  Our simulation predicted $\sim260$ detectable \lae{}s for a similar volume, quasar, and redshift.  \citet{Kashikawa:2007p5789} found that while the observed Lyman break galaxies formed a distributed filamentary structure which included the quasar, the \lae{}s were preferentially clustered around the quasar while avoiding a vicinity of $4.5\,\cmpc{}$ from the quasar.  This region was calculated to have a UV radiation field roughly $100$ times that of the background, and this was posited to be suppressing the formation of \lae{}s.  This spacial distribution of \lae{}s reinforces the idea that the environment in the smaller volumes probed by our model -- in the direct vicinity of the quasar -- is where the majority of galaxy suppression takes place.

Recent work by \citet{Laursen:2011p5732} calculated \igmtrans{} using a sophisticated hydrodynamics code that accounts for radiative transfer and interstellar resonant scattering.  Our results are consistent with the lower end of their \igmtrans{} values for corresponding redshifts, implying our semi-analytic values are conservative, and thereby reinforcing our results.

\section{Conclusions}
\label{sec:conclusions}
In this paper we have presented a semi-analytic model that predicts the number of visible \lya{} emitting galaxies around a central quasar, taking into account the quasar's impact on the local \igm{} density and neutral fraction.  The free parameters of the model are determined by fitting \lya{} and UV luminosity functions taken from large field \lya{} surveys conducted by \citetalias{Ouchi:2008p3716}.  We use this model to interpret observations of a $70\,\pmpc{}^{3}$ volume centered on the $z=2.168$ luminous quasar \jossquasar{}, in which no \lya{} emission was detected \citepalias{Francis:2004p2989}.

We find that this null detection of \lya{} emitting galaxies can only be explained in a scenario in which we introduce a simple cutoff in mass below which galaxies are unobservable. In order for our model to be consistent with observations at the $68\%$ ($90\%$) level we need to exclude all masses below at least $1.2\times10^{12}\,\msolar{}$ ($2.5\times10^{12}\,\msolar{}$), corresponding to a virial temperature greater than $3.4\times10^{6}\,\kelvin{}$.  This result may imply considerable radiative suppression of galaxy formation by the nearby quasar and motivates further observations of \lya{} emitters in the vicinity of luminous quasars.  Understanding this process in more detail will ultimately help to constrain the extent to which radiative suppression of galaxy formation took place during the epoch of reionization.

Tunable filters (TF) are a powerful approach to probing emission-line objects at any redshift \citep{Jones:2001iu}. In the coming decade, there are several facilities that are ideally suited to searches for emission-line objects, including the Osiris TF on the Grantecan 10.2m \citep{Cepa:2003kz}, MMTF on Magellan \citep{Veilleux:2010du}, and TF instruments under development for the NTT 3.5m and the SOAR 4m telescopes \citep{Marcelin:2008dn, Taylor:2010il}. All of these instruments are well adapted to studying the impact of QSOs on their environs. We envisage that IR tunable filters operating with adaptive optics will be able to push to even higher redshifts and down to lower galaxy masses.

While our simulations looked at mass cuts across the entire volume surveyed, these upcoming observations will provide the statistics required for spatial mapping of \lae{}s around their central quasar allowing for future models to constrain the critical ionizing flux required to disrupt galaxy formation.  This is a fundamental unknown required by hydrodynamical N-body simulations of galaxy formation in the vicinity of AGNs and during the reionization epoch, and can be constrained using current generation instruments.

\bibliography{main}
\end{document}